\documentclass{article}

\usepackage{arxiv}

\usepackage[utf8]{inputenc} 
\usepackage[T1]{fontenc}    
\usepackage{hyperref}       
\usepackage{url}            
\usepackage{booktabs}       
\usepackage{amsfonts}       
\usepackage{nicefrac}       
\usepackage{microtype}      
\usepackage{cleveref}       
\usepackage{lipsum}         
\usepackage{graphicx}
\usepackage[numbers]{natbib}
\usepackage{doi}
\usepackage{subcaption} 
\usepackage{xcolor}     
\pdfoutput=1 

\title{FAIR GPT: A virtual consultant for research data management in ChatGPT}

\date{September 20, 2024}

\newif\ifuniqueAffiliation
\uniqueAffiliationtrue

\ifuniqueAffiliation 
\author{ \href{https://orcid.org/0000-0002-0331-2558}{\includegraphics[scale=0.06]{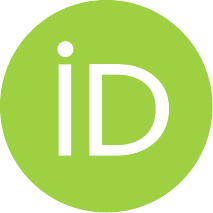}\hspace{1mm}Renat Shigapov}\thanks{\href{https://www.bib.uni-mannheim.de/ihre-ub/ansprechpersonen/dr-renat-shigapov}{https://www.bib.uni-mannheim.de/ihre-ub/ansprechpersonen/dr-renat-shigapov}} \\
	University Library\\
	University of Mannheim\\
	Germany \\
	\texttt{shigapov@uni-mannheim.de} \\
	\And
	\href{https://orcid.org/0000-0002-0167-3683}{\includegraphics[scale=0.06]{orcid.pdf}\hspace{1mm}Irene Schumm} \\
	University Library\\
	University of Mannheim\\
	Germany \\
	\texttt{schumm@uni-mannheim.de} \\
}
\else
\usepackage{authblk}

\newbox{\orcid}\sbox{\orcid}{\includegraphics[scale=0.06]{orcid.pdf}} 
\fi

\renewcommand{\shorttitle}{\textit{arXiv} Template}

\hypersetup{
pdftitle={FAIR GPT: A virtual consultant for research data management in ChatGPT},
pdfauthor={Renat Shigapov, Irene Schumm},
pdfkeywords={FAIR Principles,  FAIR Data, FAIR Research Data Management, ChatGPT},
}

\begin{document}
\maketitle

\begin{abstract}
FAIR GPT is a first virtual consultant in ChatGPT designed to help researchers and organizations make their data and metadata compliant with the FAIR (Findable, Accessible, Interoperable, Reusable) principles. It provides guidance on metadata improvement, dataset organization, and repository selection. To ensure accuracy, FAIR GPT uses external APIs to assess dataset FAIRness, retrieve controlled vocabularies, and recommend repositories, minimizing hallucination and improving precision. It also assists in creating documentation (data and software management plans, README files, and codebooks), and selecting proper licenses. This paper describes its features, applications, and limitations.
\end{abstract}

\keywords{FAIR Principles \and FAIR Data \and FAIR Research Data Management \and ChatGPT}

\section*{Introduction}
The FAIR (Findable, Accessible, Interoperable, Reusable) principles are now a widely accepted framework for scientific data management and stewardship \citep{wilkinson2016}. However, despite their importance, many researchers and organizations struggle to implement these principles effectively due to a range of cultural, technical, and organizational barriers \citep{stall2019, mons2020, jacobsen2020}.

FAIR GPT, a customized virtual consultant in ChatGPT, addresses some of these challenges by providing automated support for making data and metadata FAIR-compliant. This paper presents its features, applications by researchers and data stewards, and limitations of the tool.

\section{Features}
FAIR GPT offers features that assist researchers and data stewards in various aspects of research data management (RDM). These features help users ensure their data complies with the FAIR principles, and meet the needs of their institutions and funders.

\begin{itemize}
\item \textbf{RDM consultancy.} FAIR GPT acts as a virtual consultant, providing guidance on best practices in research data management. It assists researchers and organizations in structuring their data workflows, ensuring compliance with FAIR principles, and addressing challenges related to data stewardship. 
\item \textbf{Metadata review.} It reviews uploaded or copy-pasted metadata, assessing it against international standards. The tool provides suggestions for improving metadata, ensuring compliance with FAIR principles. Additionally, FAIR GPT connects to external resources such as the \href{https://terminology.tib.eu/ts/api}{TIB Terminology Service API} to suggest terms from controlled vocabularies and ontologies, and to the \href{https://www.wikidata.org/w/api.php}{Wikidata API} to ensure the correct use of Wikidata identifiers. Without querying external terminology services, ChatGPT may generate inaccurate or fictitious identifiers for terms from controlled vocabularies, leading to potential errors in metadata and vocabulary alignment \citep{shigapov_2023, shigapov_2024}."
\item \textbf{Data organization.} Organizing datasets properly is critical for their reusability. FAIR GPT recommends best practices for dataset organization, offering guidance on optimal folder structures, file naming conventions, and data hierarchies.
\item \textbf{Documentation creation.} FAIR GPT assists users in generating key documentation necessary for proper data stewardship. These include Data Management Plans (DMPs), Software Management Plans (SMPs), README files, and codebooks. Based on the provided dataset and code, FAIR GPT tailors the documentation to the specific needs of the project, ensuring that the data can be easily understood, reused, and cited.
\item \textbf{FAIR assessment.} It supports two open APIs for FAIR data assessment: FAIR-Checker \cite{gaignard2023} and FAIR-Enough \cite{fair_enough}. By analyzing how well a dataset adheres to FAIR principles, FAIR GPT provides actionable recommendations for improving findability, accessibility, interoperability, and reusability. This helps users enhance the overall quality of their data and metadata.
\item \textbf{Data Licensing Recommendation.} FAIR GPT provides guidance on selecting appropriate data licenses based on the type of dataset, its intended use, and the relevant legal and institutional frameworks. This ensures that data sharing is legally compliant while promoting reuse.
\item \textbf{Data repository selection.} Selecting an appropriate repository is crucial for ensuring long-term data archiving and compliance with FAIR principles. FAIR GPT leverages the \href{https://www.re3data.org/api/beta}{re3data API} to recommend suitable repositories.
\item \textbf{Data paper publication.} FAIR GPT also assists researchers in identifying suitable data journals for publishing data papers, which increases the visibility of their datasets and enhances their citation potential. It provides recommendations for journals such as Scientific Data, Data in Brief, and other relevant outlets based on the research domain and our instructions\footnote{\href{https://github.com/UB-Mannheim/FAIR-GPT/blob/main/GPT/Instructions.md}{https://github.com/UB-Mannheim/FAIR-GPT/blob/main/GPT/Instructions.md}}.
\end{itemize}

FAIR GPT also utilizes the uploaded research data management resources, including the "Guidelines on FAIR Data Management in Horizon 2020" \cite{H2020Programme}, "Turning FAIR into reality – Final report and action plan from the European Commission expert group on FAIR data" \cite{turningFAIR}, and "A curated GitHub-list of awesome RDM resources for researchers and organizations" \cite{awesomeRDM}. These resources help FAIR GPT to provide well-informed recommendations for RDM best practices.

The FAIR GPT assets (frontend, documentation, instructions, and issues) with their URLs are described in Table~\ref{tab:assets}. Suggestions for further improvements are collected via GitHub issues. Instructions for FAIR GPT are openly shared at GitHub. Fürther documentation is also openly available at GitHub.

\begin{table}[ht]
    \caption{FAIR GPT assets}
    \centering
    \begin{tabular}{ll}
        \toprule
        Asset  & URL \\
        \midrule
        Frontend & \href{https://chat.openai.com/g/g-BkMR28wlV-fair}{https://chat.openai.com/g/g-BkMR28wlV-fair} \\
        Documentation & \href{https://github.com/UB-Mannheim/FAIR-GPT}{https://github.com/UB-Mannheim/FAIR-GPT} \\
        Instructions & \href{https://github.com/UB-Mannheim/FAIR-GPT/blob/main/GPT/Instructions.md}{https://github.com/UB-Mannheim/FAIR-GPT/blob/main/GPT/Instructions.md} \\
        Issues & \href{https://github.com/UB-Mannheim/FAIR-GPT/issues}{https://github.com/UB-Mannheim/FAIR-GPT/issues} \\
        \bottomrule
    \end{tabular}
    \label{tab:assets}
\end{table}

\section{Applications}

\begin{figure}[ht]
    \centering
    \begin{subfigure}[b]{0.49\textwidth}
        \centering
        \fboxsep=2mm
        \fcolorbox{gray}{white}{\includegraphics[width=\textwidth]{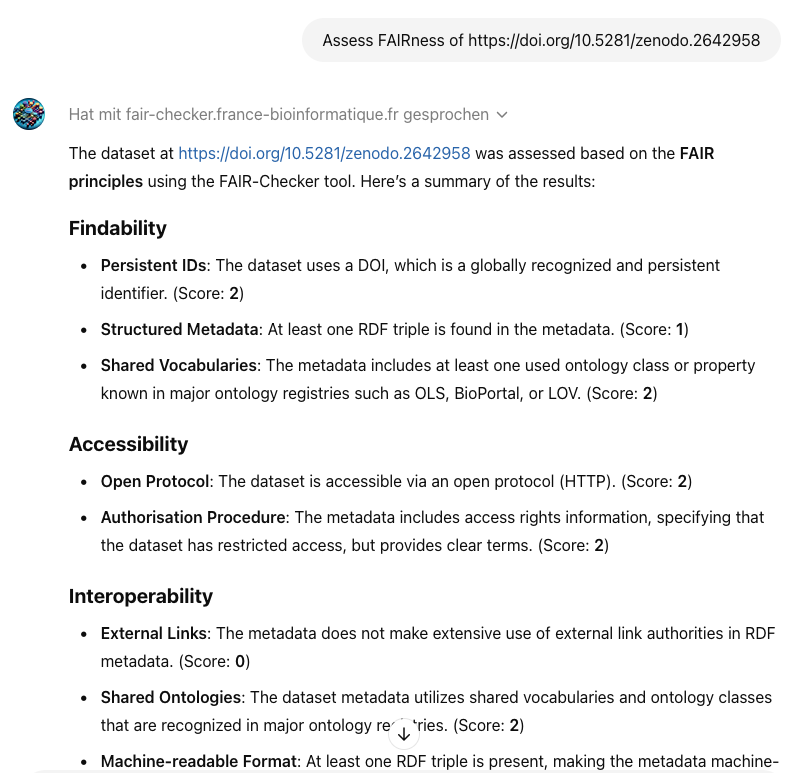}}
        \caption{A researcher asks to assess FAIRness of data.}
        \label{fig:fair_gpt_for_researchers}
    \end{subfigure}
    \hfill
    \begin{subfigure}[b]{0.46\textwidth}
        \centering
        \fboxsep=2mm
        \fcolorbox{gray}{white}{\includegraphics[width=\textwidth]{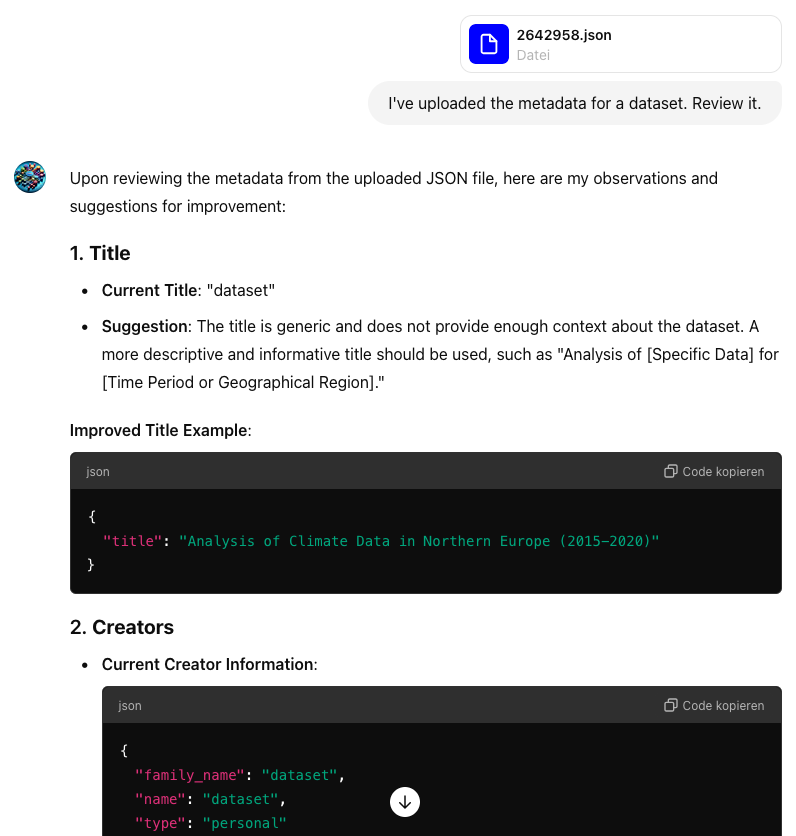}}
        \caption{A data steward asks to review metadata.}
        \label{fig:fair_gpt_for_data_stewards}
    \end{subfigure}
    \caption{FAIR GPT for different roles in research data management.}
    \label{fig:fair_gpt_combined}
\end{figure}

FAIR GPT supports both researchers and data stewards by automating key aspects of research data management, thus reducing manual effort and improving the overall quality and FAIRness of datasets (see Figure~\ref{fig:fair_gpt_combined}).

\textbf{For researchers.} FAIR GPT simplifies the process of preparing data and metadata for sharing and publication. Researchers can use the tool to draft rich metadata, generate essential documentation (e.g., README files, codebooks, DMPs, and SMPs), and select suitable data repositories for long-term archiving. By automating these tasks, FAIR GPT allows researchers to focus on their core scientific activities while ensuring their datasets meet the FAIR principles. Once a dataset is published, researchers can also ask FAIR GPT to assess its FAIRness by providing the dataset’s DOI, as illustrated in Figure~\ref{fig:fair_gpt_for_researchers}. This feature helps researchers check whether the data is FAIR after publication.

\textbf{For data stewards.} FAIR GPT can assist data stewards with reviewing and improving datasets submitted to data repositories. The tool evaluates the quality of metadata, assesses the organization of datasets, and reviews accompanying documentation to ensure compliance with FAIR standards. At the Research Data Center of Mannheim University Library, for example, we leverage FAIR GPT to automatically generate initial dataset reviews (see Figure~\ref{fig:fair_gpt_for_data_stewards}). These reviews are then fine-tuned by data stewards before being shared with researchers, streamlining the review process and helping to ensure that deposited datasets meet institutional and FAIR criteria. By integrating FAIR GPT into their workflows, data stewards can improve the quality and FAIRness of repository submissions.

\section{Limitations}

While FAIR GPT offers useful features, it also has several limitations that affect its applicability in real-world practice:

\begin{itemize}
\item \textbf{Hallucinations.} Despite using external APIs to minimize hallucinations, FAIR GPT may still generate incorrect or misleading recommendations, especially in novel or ambiguous cases. These hallucinations can introduce errors into metadata recommendations, reducing the overall reliability of the tool’s outputs.
\item \textbf{Lack of provenance for generated data.} FAIR GPT does not provide clear sources for recommendations it generates. This lack of transparency makes it difficult for users to trace the origins of suggested improvements, reducing the trust in the generated answers.
\item \textbf{Evolving data management practices.} The field of research data management is continually evolving, and FAIR GPT requires continuous updates to stay relevant. Without frequent updates, the tool risks becoming outdated and ineffective.
\item \textbf{Privacy concerns for sensitive data.} FAIR GPT is not specifically designed to handle sensitive or restricted data, and it may not adequately account for legal or institutional protocols related to personal data, such as anonymization and compliance with data protection regulations (e.g., GDPR). Users dealing with sensitive datasets should refrain from uploading or copy-pasting such data into FAIR GPT, as it may introduce privacy risks.
\item \textbf{There is no API.} FAIR GPT does not offer an API for external integration, which limits its ability to be embedded into automated workflows or custom research data management systems. As a result, users must rely on manual interaction through the graphical interface, which can be time-consuming and less efficient.
\end{itemize}

\section*{Conclusions}

FAIR GPT is the first virtual consultant designed to assist with research data management in compliance with the FAIR principles. It automates tasks such as metadata enhancement, dataset organization, and documentation creation, reducing the manual effort required for FAIR compliance. By integrating external APIs, FAIR GPT improves the precision of its recommendations. However, the tool has limitations, including potential hallucinations, lack of provenance for generated data, and no API for external integration, which limits scalability. Privacy concerns restrict its use with sensitive datasets. Despite these limitations, FAIR GPT offers valuable support by streamlining data management tasks and improving metadata quality. Addressing challenges such as hallucinations and lack of provenance could further enhance its applicability.

\section*{Funding}

This work has been partially funded by the Deutsche Forschungsgemeinschaft (DFG, German Research Foundation) as part of BERD@NFDI with grant number 460037581 (\href{https://gepris.dfg.de/gepris/projekt/460037581}{https://gepris.dfg.de/gepris/projekt/460037581}).

\section*{Conflict of interest}

The authors have no conflict of interest to report.

\bibliographystyle{unsrtnat}
\bibliography{references}  

\end{document}